\documentclass[12pt,a4paper]{article}

 \newcommand{\lyxaddress}[1]{
   \par {\raggedright #1 
   \vspace{1.4em}
   \noindent\par}
 }

\usepackage{amsmath,amsfonts,amsthm,amssymb}
\usepackage{mathrsfs}
\usepackage{graphicx}

\theoremstyle{remark}
\newtheorem*{rem*}{Remark}
\newtheorem*{rems*}{Remarks}
\theoremstyle{definition}

\newtheorem*{def*}{Definition}

\newcommand{\R}{\mathbb{R}}
\newcommand{\Z}{\mathbb{Z}}

\newcommand{\dd}{\mathrm{d}}
\newcommand{\const}{\mathrm{const}}

\begin{document}

\title{On the harmonic oscillator on the Lobachevsky plane}

\date{{}}

\author{P.~\v{S}\v{t}ov\'\i\v{c}ek, M.~Tu\v{s}ek}

\maketitle

\lyxaddress{Department of Mathematics, Faculty of Nuclear
  Science, Czech Technical University, Trojanova 13, 120 00 Prague,
  Czech Republic}

\begin{abstract}
  \noindent
  We introduce the harmonic oscillator on the Lobachevsky plane with
  the aid of the potential
  $V(\varrho)=(a^{2}\omega^{2}/4)\sinh(\varrho/a)^{2}$ where $a$ is
  the curvature radius and $\varrho$ is the geodesic distance from a
  fixed center. Thus the potential is rotationally symmetric and
  unbounded likewise as in the Euclidean case. The eigenvalue equation
  leads to the differential equation of spheroidal functions.  We
  provide a basic numerical analysis of eigenvalues and eigenfunctions
  in the case when the value of the angular momentum, $m$, equals $0$.
\end{abstract}

\section{Introduction}

This short contribution is dedicated to the memory of Vladimir Geyler,
our dear colleague and friend who, to our great sorrow, passed
recently and very unexpectedly away. Among his numerous scientific
activities in the framework of mathematical physics one can point out
his interest in the influence of nontrivial geometries on the
properties of quantum mechanical systems which was very intensive
especially in the last period. Let us mention just a few of his
results focused on this sort of problems: the effect of surface
curvature on magnetic moment and persistent currents \cite{bgm}, the
general form of on-diagonal singularities of the Green function for
Schr\"{o}dinger operators on curved spaces \cite{ondiag}, and zero
modes in a system of Aharonov-Bohm solenoids on the Lobachevsky plane
\cite{gs}.
%

In one of the very last scientific contributions due to Vladimir
Geyler \cite{gst}, on which we had the pleasure to collaborate, a
quantum dot with impurity in the Lobachevsky plane is discussed. This
paper was intended as a step towards an extension to the hyperbolic
plane of results derived in \cite{bgl} where a similar problem is
treated in the case of the Euclidean geometry. The current article
aims to complete the discussion from \cite{gst} by a short numerical
analysis of spectral properties of the harmonic oscillator on the
Lobachevsky plane. In particular, we are interested in the effect of
the surface curvature on these properties.

It has been observed already some time ago that various solvable
quantum mechanical systems admit an explicit treatment also in the
case of hyperbolic geometry, though with interesting nontrivial
modifications (see, for example, \cite{com,aco}). Also the spectral
problem for the quantum harmonic oscillator on the Lobachevsky plane,
as we introduce it below, leads to a differential equation which is
well known from the theory of special functions, namely to the
differential equation of spheroidal functions. It should be
emphasized, however, that the history of spheroidal functions is much
more recent than that of more traditional special functions like
Bessel functions or Legendre polynomials. For example, one of the
basic monographs devoted to these functions appeared only in the
fifties of the last century \cite{me}. Also the notation is still not
fully uniform. One can compare, for example, \cite{me} with \cite{be}.
Here we follow the latter source. Furthermore, there are cases of
values of parameters on which the spheroidal functions depend that
have not been fully investigated. In this connection we note that
particularly the values of parameters which are of interest for our
model are treated in the textbooks in a rather marginal way. Also
effective numerical algorithms to evaluate the spheroidal functions
seem to be rather tedious to create and not available for all cases.

These circumstances make the numerical and qualitative analysis of the
spectrum more complicated than one might expect at first glance.  The
numerical results presented in this contribution were derived with the
aid of the computer algebra system \textit{Mathematica}. On the
contrary to other ``more traditional'' special functions, the
spheroidal functions were implemented in this system only in the
version which was the very last one at the time of this writing.

\section{The Hamiltonian}

Denote by $a$, $0<a<\infty$, the so called curvature radius which is
related to the scalar curvature by the formula $R=-2/a^2$, and by
$(\varrho,\phi)$, $0<\varrho<\infty$, $0\leq\phi<2\pi$, the geodesic
polar coordinates on the Lobachevsky plane. Taking the mass equal
$1/2$ and $\hbar=1$, the Hamiltonian has the form
\begin{displaymath}
  H = -\left(\frac{\partial^2}{\partial\varrho^2}
    +\frac{1}{a}\coth\!\left(\frac{\varrho}{a}\right)
    \frac{\partial}{\partial\varrho}
    +\frac{1}{a^2\sinh\!\left(\frac{\varrho}{a}\right)^2}\,
    \frac{\partial^2}{\partial\phi^2}+\frac{1}{4a^2}\right)
  +V(\varrho),
\end{displaymath}
and is defined in
$L^2((0,\infty)\times{}S^1,a\sinh(\varrho/a)\dd\varrho\,\dd\phi)$ on
its natural domain. Here $V(\varrho)$ is the potential of the harmonic
oscillator on the hyperbolic plane. Its choice is ambiguous. Of
course, one requires that in the flat limit, $a\to\infty$, $V(\rho)$
becomes the usual potential of the isotropic harmonic oscillator on
the Euclidean plane, $V^\infty(\varrho)=\omega^2\varrho^2/4$. At least
three choices of $V(\varrho)$ have been discussed in the literature so
far.

A distinguished possibility is
\begin{displaymath}
  V_1(\varrho) = \frac{1}{4}\,a^{2}\omega^{2}
  \tanh\!\left(\frac{\varrho}{a}\right)^{\!2}.
\end{displaymath}
since in that case one gets a Hamiltonian with a comparatively rich
symmetry, a so called super-integrable model. this is demonstrated in
detail on the level of classical mechanics in \cite{ran}, see also
\cite{ranII}. Notice that this potential is bounded and thus the
spectrum of the corresponding quantum Hamiltonian contains an
absolutely continuous part, the interval $[E_0(a),\infty[\,$. The 
spectral analysis for this case is covered by paper \cite{bgm}. It
turns out that $E_0(a)\sim{}a^2$ as $a\to\infty$, and so the
continuous part disappears in the flat limit. Moreover, as expected,
the number of eigenvalues in the curved case is finite but tends to
infinity in the flat limit, and the eigenvalues approach the discrete
spectrum of the flat Hamiltonian.

Further, a nonlinear quantum model with a potential of the type
\begin{displaymath}
  V_2(\varrho) = \const\,\frac{\varrho^2}{1+\lambda\varrho^2}
\end{displaymath}
depending on a parameter $\lambda$ is discussed in \cite{crs}. The
treatment of this model also relies heavily on symmetry properties
which allow for explicit solutions.

In the current paper we stick, however, to a choice made in \cite{gst}
and otherwise not discussed in the above cited papers. We set
\begin{displaymath}
  V(\varrho) = \frac{1}{4}\,a^{2}\omega^{2}
  \sinh\!\left(\frac{\varrho}{a}\right)^{\!2}.
\end{displaymath}
With this choice, the potential is unbounded, likewise as in the
Euclidean case. Consequently, the corresponding Hamiltonian $H$ has a
purely discrete spectrum depending on the parameter $a$.

Next we rescale the Hamiltonian letting
\begin{displaymath}
  \tilde{H} = a^2H.
\end{displaymath}
Substituting $\xi=\cosh(\varrho/a)$ we obtain
\begin{displaymath}
  \tilde{H} = (1-\xi^{2})\,\frac{\partial^2}{\partial\xi^{2}}
  -2\xi\,\frac{\partial}{\partial\xi}
  +(1-\xi^2)^{-1}\frac{\partial^{2}}{\partial\phi^{2}}
  + \frac{a^{4}\omega^{2}}{4}\,(\xi^{2}-1)-\frac{1}{4}
\end{displaymath}
in $L^2((1,\infty)\times{}S^1,\dd\xi\,\dd\phi)$. Using the rotational
symmetry which amounts to a Fourier transform in the variable $\phi$,
$\tilde{H}$ may be decomposed into a direct sum as follows
\begin{eqnarray*}
  && \tilde{H}
  = \sideset{}{^\oplus}\sum_{m=-\infty}^\infty\,\tilde{H}_m, \\
  && \tilde{H}_{m}
  = -\frac{\partial}{\partial\xi}
  \left((\xi^{2}-1)\frac{\partial}{\partial\xi}\right)
  +\frac{m^2}{\xi^2-1}+\frac{a^{4}\omega^{2}}{4}\,(\xi^{2}-1)
  -\frac{1}{4}\,.
\end{eqnarray*}
For all $m\in\Z$, $\tilde{H}_m$ acts in $L^2((1,\infty),\dd\xi)$.

\section{Eigenvalues and eigenfunctions}

Denote by $H_m$, $m\in\Z$, the restrictions of $H$ to the eigenspace
of the angular momentum with value $m$. We prefer to regard $H_m$ as
an operator in $L^2(\R_+,\dd\varrho)$. This assumes an obvious unitary
transform to pass from the measure $a\sinh(\varrho/a)\dd\varrho$ in
the $L^2$ space on $\R_+$ to the measure $\dd\varrho$. On the general
grounds, the spectrum of $H_m$ is semibounded below, discrete and
simple \cite{bs,weidman}. Set
\begin{displaymath}
  q = \frac{a^2\omega}{2}\,.
\end{displaymath}
If $\tilde{E}^m_n$, $n=0,1,2,\ldots$, is the $n$the eigenvalue of
$\tilde{H}_m$ then
\begin{displaymath}
  E^m_n = \frac{1}{a^2}\,\tilde{E}^m_n
  = \frac{\tilde{E}^m_n}{2q}\,\omega
\end{displaymath}
is the $n$th eigenvalue of $H_m$.

The eigenvalue equation $\tilde{H}_m\psi=E{}\psi$ written in the
coordinate $\xi$ takes the form
\begin{equation}
  \label{eq:spheroidal}
  (1-\xi^{2})\frac{\partial^2\psi}{\partial\xi^{2}}
  -2\xi\frac{\partial\psi}{\partial\xi}
  +\left(\lambda+4\theta(1-\xi^2)-\frac{m^2}{1-\xi^2}\right)\psi
  = 0
\end{equation}
where
\begin{displaymath}
  \lambda = -E-\frac{1}{4},\textrm{~}4\theta
  =-\frac{a^2\omega^2}{4}=-q^2.
\end{displaymath}
Differential equation (\ref{eq:spheroidal}) is known in the theory of
special functions as the differential equation of spheroidal functions
\cite{be,me}. The notation used below follows the source \cite{be}.
All parameters in (\ref{eq:spheroidal}) are in general complex
numbers. There are two solutions that behave like $\xi^{\nu}$ times a
single-valued function and $\xi^{-\nu-1}$ times a single-valued
function at $\infty$. The exponent $\nu$ is a function of $\lambda$,
$\theta$, $m$, and is called the characteristic exponent. Usually, it
is more convenient to regard $\lambda$ as a function of $\nu$, $m$ and
$\theta$. We shall write $\lambda=\lambda^{m}_{\nu}(\theta)$. The
functions $\lambda^{m}_{\nu}(\theta)$ obey the symmetry relations
\begin{equation}
  \label{eq:lambdasym}
  \lambda^{m}_{\nu}(\theta)
  = \lambda^{-m}_{\nu}(\theta)
  = \lambda^{m}_{-\nu-1}(\theta) 
  = \lambda^{-m}_{-\nu-1}(\theta).
\end{equation}

A first group of solutions (the so called radial spheroidal functions)
is obtained as expansions in series of Bessel functions. They are
denoted $S^{m(j)}_{\nu}(\xi,\theta)$, $j=1,2,3,4$. Several relations
between the radial spheroidal functions are known, notably
\begin{displaymath}
  S^{m(3)}_{\nu}(\xi,\theta)
  = S^{m(1)}_{\nu}(\xi,\theta)+i\,S^{m(2)}_{\nu}(\xi,\theta)
  = -i\,e^{-i\nu\pi}S^{m(3)}_{-\nu-1}(\xi,\theta)
\end{displaymath}
and
\begin{displaymath}
  S^{m(3)}_{\nu}(\xi,\theta) = \frac{1}{i\cos(\nu\pi)}
  \left(S^{m(1)}_{-\nu-1}(\xi,\theta)
    +i\,e^{-i\nu\pi}S^{m(1)}_{\nu}(\xi,\theta)\right).
\end{displaymath}
Also the asymptotic behavior of radial spheroidal functions is well
known. It turns out that among these functions the only one which is
square integrable near infinity is $S^{m(3)}_{\nu}(\xi,\theta)$. For
$q>0$ it holds
\begin{displaymath}
  S^{m(3)}_{\nu}\!\left(\xi,-\frac{q^2}{4}\right)
  = -\frac{1}{q\xi}\,\exp\!\left(-q\xi-i\,\frac{\nu\pi}{2}\right)
  \left(1+O(\xi^{-1})\right)
\end{displaymath}
as $\xi\to+\infty$.

To control the behavior near the singular point $\xi=1$ one uses
expansions in series in Legendre functions. These solutions are called
angular spheroidal functions and are denoted $Ps^m_{\nu}(\xi,\theta)$
and $Qs^m_{\nu}(\xi,\theta)$. There exist relations between these two
types of spheroidal functions. In particular, it holds
\begin{displaymath}
  S^{m(1)}_{\nu}(\xi,\theta)
  = V^m_\nu(\theta)Qs^m_{-\nu-1}(\xi,\theta)
\end{displaymath}
where $V^m_\nu(\theta)$ is a coefficient independent of $\xi$.
Consequently,
\begin{displaymath}
  S^{m(3)}_{\nu}(\xi,\theta) = \frac{1}{i\cos(\nu\pi)}
  \left(V^m_{-\nu-1}(\theta)Qs^m_{\nu}(\xi,\theta)
    +i\,e^{-i\nu\pi}V^m_\nu(\theta)Qs^m_{-\nu-1}(\xi,\theta)\right).
\end{displaymath}

The angular spheroidal functions asymptotically behave, as
$\xi\to+\infty$, in the following way:
\begin{displaymath}
  Ps^m_{\nu}(\xi,\theta) = \frac{1}{\Gamma(1-m)s^m_\nu(\theta)}
  \left(\frac{\xi-1}{2}\right)^{\!-m/2}\left(1+O(\xi-1)\right)
\end{displaymath}
where $s^m_\nu(\theta)$ is some coefficient (for its definition we
refer to \cite{be}). Particularly, we have
\begin{displaymath}  
  \frac{1}{s^0_\nu(\theta)} = Ps^0_{\nu}(1,\theta).
\end{displaymath}
Furthermore,
\begin{displaymath}
  Qs^0_{\nu}(\xi,\theta) = -\frac{1}{2s^0_\nu(\theta)}
  \log(\xi-1)+O(1)
\end{displaymath}
and
\begin{displaymath}
  (\xi-1)\,\frac{\partial Qs^0_{\nu}(\xi,\theta)}{\partial\xi}
  = -\frac{1}{2s^0_\nu(\theta)}+o(1).
\end{displaymath}
It follows that
\begin{displaymath}
  S^{0(3)}_{\nu}(\xi,\theta) = -\frac{1}{2i\cos(\nu\pi)}
  \left(\frac{V^0_{-\nu-1}(\theta)}{s^0_\nu(\theta)}
    +i\,e^{-i\nu\pi}\frac{V^0_\nu(\theta)}{s^0_{-\nu-1}(\theta)}\right)
  \log(\xi-1)+O(1).
\end{displaymath}

We conclude that $\tilde{\psi}(\xi)=S^{m(3)}_{\nu}(\xi,-q^2/4)$ is an
eigenfunction and, at the same time,
$\tilde{E}^m_\nu=-\lambda^m_\nu(-q^2/4)-1/4$ is an eigenvalue of
$\tilde{H}_m$ if and only if $\tilde{\psi}(\xi)$ has no singularity as
$\xi\to1+$. Further we restrict ourselves to the case $m=0$. The
function $\lambda^0_\nu(\theta)$ is real for $\nu=-1/2+it$, $t\in\R$,
and becomes negative if $t$ is sufficiently large. The eigenvalues of
$\tilde{H}_0$ are the numbers $-\lambda^0_{-1/2+it}(-q^2/4)-1/4$ where
$t$ solves the equation
\begin{displaymath}
  e^{-t\pi/2}\frac{V^0_{-1/2-it}(-q^2/4)}{s^0_{-1/2+it}(-q^2/4)}
  -e^{t\pi/2}\frac{V^0_{-1/2+it}(-q^2/4)}{s^0_{-1/2-it}(-q^2/4)}
  = 0.
\end{displaymath}
Because of symmetry (\ref{eq:lambdasym}) it suffices to consider
$t>0$. Equivalently, in the numerical computations the following form
of this equation proved itself to be rather effective:
\begin{displaymath}
  \lim_{\xi\to1+}\,(\xi-1)\,
  \frac{\partial S^{0(3)}_{-1/2+it}(\xi,-q^2/4)}{\partial\xi}
  = 0.
\end{displaymath}

Finally, we present a few numerical results. Figure~\ref{fig:evs}
depicts several first eigenvalues of $H_0$. More precisely, the figure
contains plots of eigenvalues $E^0_n(q)/\omega=\tilde{E}^0_n(q)/(2q)$
for $n=0,1,2$. These functions of the variable $q=a^2\omega/2$
decrease monotonically, and in the limit $q\to\infty$ they approach
the eigenvalues in the flat case, namely $2n+1$. On the other hand,
for $q\to0+$ the eigenvalues tend to infinity.

Next we plot the corresponding eigenfunctions for several values of
the parameter $q$. We have chosen $q=0.5,5,\infty$. The last value
corresponds to the flat case. To reach this goal it was necessary to
fix the value of $\omega$. We have taken $\omega=1$. Then the chosen
values of $q$ correspond to the curvature radius $a^2=1,10,\infty$.
An eigenfunction $\tilde{\psi}(\xi)$ of $\tilde{H}_0$ equals
$S^{0(3)}_{\nu}(\xi,-q^2/4)$ times a normalization factor for a
suitable value of $\nu\in-1/2+i\R_+$. The plotted eigenfunctions are
expressed in the variable $\varrho$. Thus the transformed
eigenfunction of $H_0$ in $L^2(\R_+,\dd\varrho)$ equals
\begin{displaymath}
  \psi(\varrho) = \left(\frac{1}{a}\,
    \sinh\!\left(\frac{\varrho}{a}\right)\right)^{\!1/2}
    \tilde{\psi}
    \!\left(\cosh\!\left(\frac{\varrho}{a}\right)\right).
\end{displaymath}
The fact that we stick to the same Hilbert space in all cases, namely
to $L^2(\R_+,\dd\varrho)$, facilitates the comparison of
eigenfunctions for various values of the curvature radius. The same
Hilbert space is used also in the limiting Euclidean case when
$a=\infty$. The eigenfunctions for the flat case are, of course, well
known from textbooks. Let us recall that in the Euclidean plane we
have
\begin{displaymath}
  \psi^0_n(\varrho) = (-1)^n\sqrt{\varrho}\,
  L^0_n\!\left(\frac{\varrho^2}{2}\right)
  \exp\!\left(-\frac{\varrho^2}{4}\right).
\end{displaymath}
Looking at figures \ref{fig:ef0}, \ref{fig:ef1} and \ref{fig:ef2}
depicting the eigenfunctions indexed $n=0$, $1$ and $2$, respectively,
one observes clearly that a more curved space (corresponding to
smaller $a$) forces the eigenfunctions to concentrate more closely to
the center of the harmonic oscillator.

\subsection*{Acknowledgments}

One of the authors (P.~\v{S}.) wishes to acknowledge gratefully a
partial support from grant No. 201/05/0857 of the Grant Agency of
Czech Republic.

\begin{figure}[hp]
  {\centering \includegraphics[width=12cm]{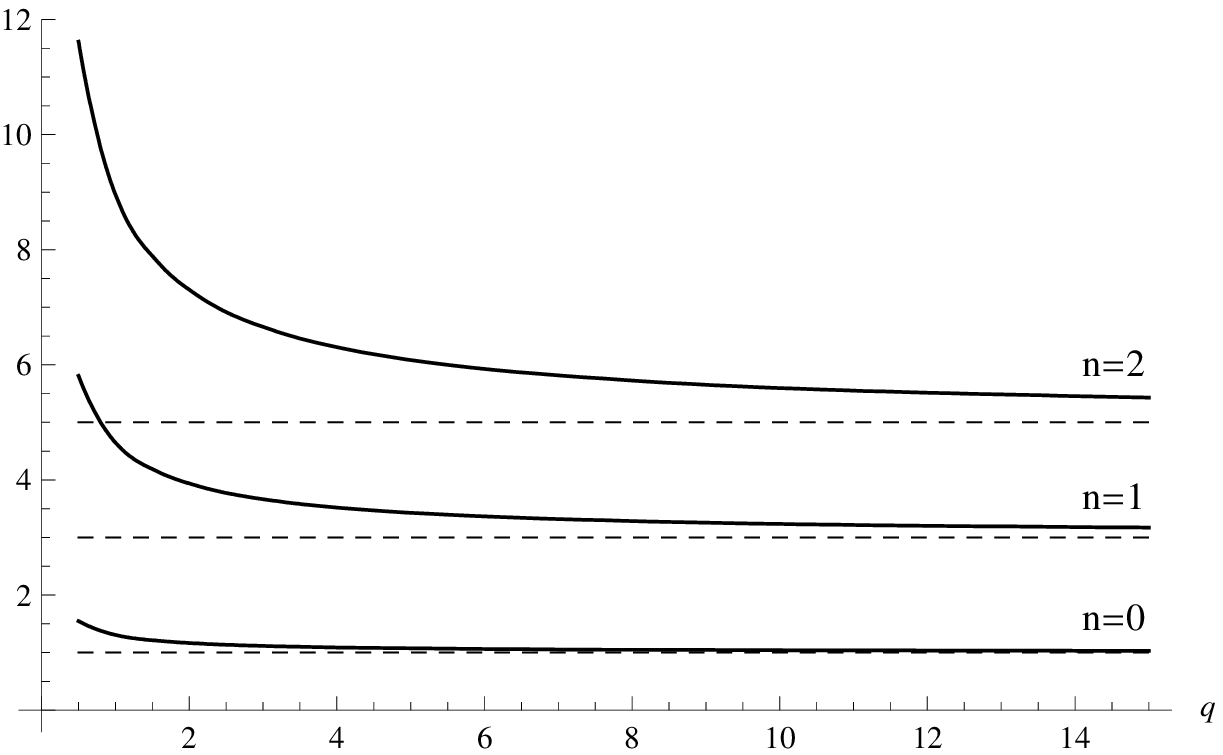} \par}
  \vskip 116pt
  \caption{}
  \label{fig:evs}
\end{figure}

\newpage
\begin{center}
\textbf{\Large Figure captions}
\end{center}
\vskip 12pt
\noindent
FIGURE~\ref{fig:evs}. Plots of
$E^0_n(q)/\omega=\tilde{E}^0_n(q)/(2q)$, $n=0,1,2$, where $E^0_n$ is
the $n$th eigenvalue of $H_0$ and $q=a^2\omega/2$.
\newline\vskip 24pt

\begin{figure}[hp]
  {\centering \includegraphics[width=12cm]{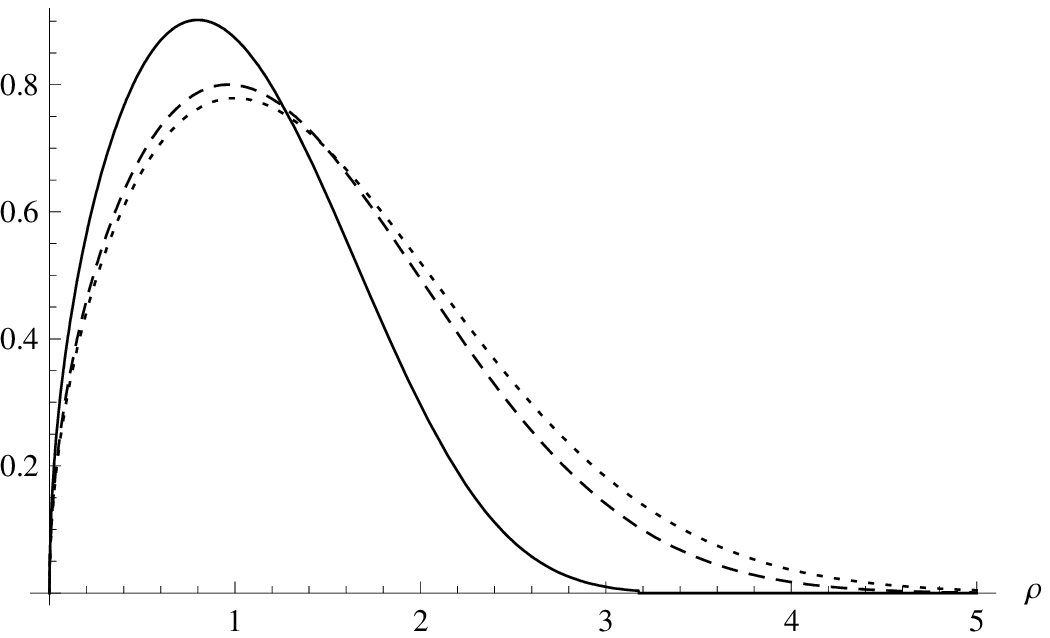}\par}
  \vskip 116pt
  \caption{}
  \label{fig:ef0}
\end{figure}

\begin{figure}[hp]
  {\centering \includegraphics[width=12cm]{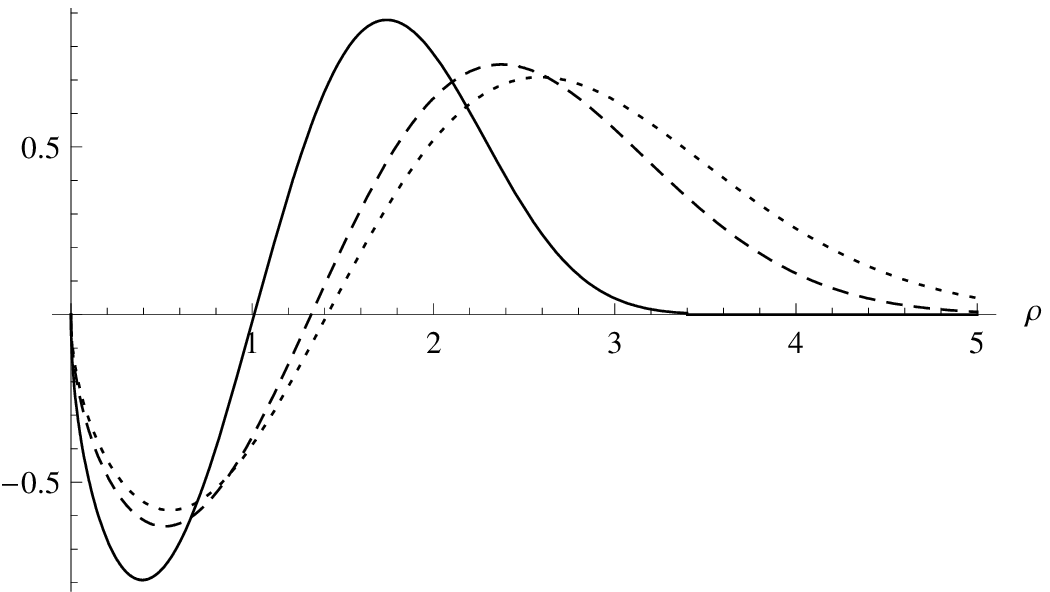}\par}
  \vskip 116pt
  \caption{}
  \label{fig:ef1}
\end{figure}

\begin{figure}[hp]
  {\centering \includegraphics[width=12cm]{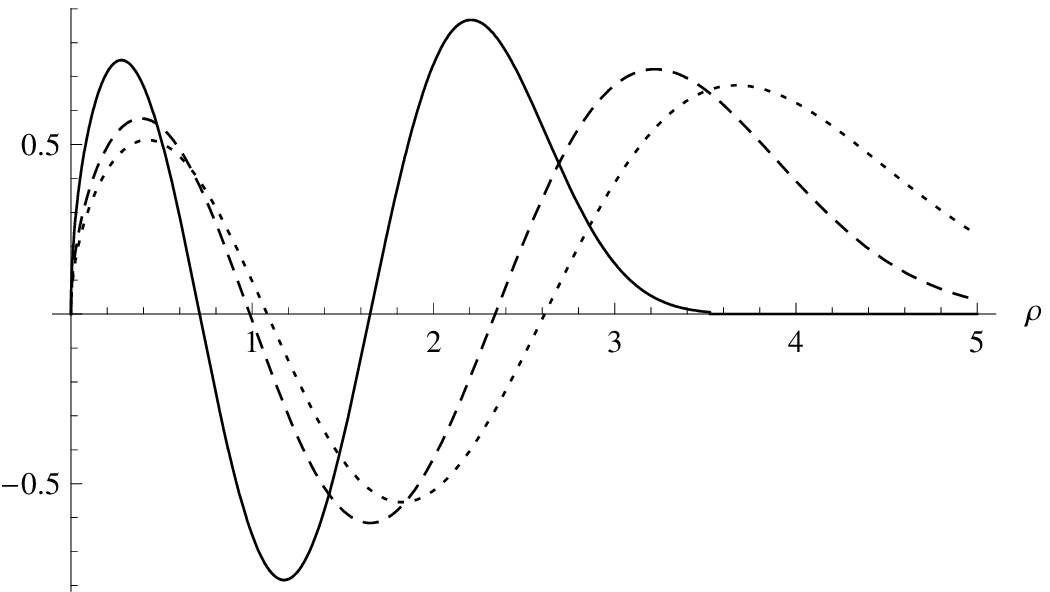}\par}
  \vskip 116pt
  \caption{}
  \label{fig:ef2}
\end{figure}

\noindent
FIGURE~\ref{fig:ef0}. The normalized eigenfunction of $H_0$,
$\psi^0_0(\varrho)\in{}L^2(\R_+,\dd\varrho)$, for the values
$\omega=1$ and $a^2=1$ (solid), $a^2=10$ (dashed), $a^2=\infty$
($=$~the flat case, dotted).
\newline\vskip 24pt

\noindent
FIGURE 3. The normalized eigenfunction of $H_0$,
$\psi^0_1(\varrho)\in{}L^2(\R_+,\dd\varrho)$, for the values
$\omega=1$ and $a^2=1$ (solid), $a^2=10$ (dashed), $a^2=\infty$
($=$~the flat case, dotted).
\newline\vskip 24pt

\noindent
FIGURE 4. The normalized eigenfunction of $H_0$,
$\psi^0_2(\varrho)\in{}L^2(\R_+,\dd\varrho)$, for the values
$\omega=1$ and $a^2=1$ (solid), $a^2=10$ (dashed) and $a^2=\infty$
($=$~the flat case, dotted).

\newpage

\newpage

\newpage

\newpage

\end{document}